\documentclass[preprint,12pt]{elsarticle}

\usepackage{graphicx}
\usepackage{amssymb}

\usepackage{gensymb}

\journal{Materials Today Chemistry}

\makeatletter
\def\ps@pprintTitle{%
  \let\@oddhead\@empty
  \let\@evenhead\@empty
  \let\@oddfoot\@empty
  \let\@evenfoot\@oddfoot
}
\makeatother

\begin{document}
\begin{frontmatter}

\title{Importance of the catalytic effect of the substrate in the functionality of lubricant additives: the case of MoDTC}



\author[Bologna]{Stefan Peeters}
\ead{stefan.peeters@unibo.it}

\author[Total]{Catherine Charrin}

\author[Total]{Isabelle Duron}

\author[Total]{Sophie Loehl\'e}

\author[Total]{Benoit Thiebaut}

\author[Bologna]{M. C. Righi\corref{1}}
\ead{clelia.righi@unibo.it}
\cortext[1]{Corresponding author}

\address[Bologna]{Department of Physics and Astronomy, University of Bologna, 40127 Bologna, Italy}

\address[Total]{Total Marketing and Services, Chemin du Canal BP 22, 69360 Solaize, France}

\begin{abstract}
Molybdenum dithiocarbamates (MoDTCs) are lubricant additives very efficient in reducing the friction of steel, and they are used in a number of industrial applications. The functionality of these additives is ruled by the chemical interactions occurring at the buried sliding interface, which are of key importance for the improvement of the lubrication performance. Yet, these tribochemical processes are very difficult to monitor in real time. Ab initio molecular dynamics simulations are the ideal tool to shed light on such a complicated reactivity. In this work, we perform ab initio simulations, both in static and tribological conditions, to understand the effect of surface oxidation on the tribochemical reactivity of MoDTC, and we find that when the surfaces are covered by oxygen, the first dissociative steps of the additives are significantly hindered. Our preliminary tribological tests on oxidized steel discs support these results. Bare metallic surfaces are necessary for a stable adsorption of the additives, their quick decomposition, and the formation of a durable MoS$_2$ tribolayer. This work demonstrates the importance of the catalytic role of the substrate and confirms the full capability of the computational protocol in the pursuit of materials and compounds more efficient in reducing friction.
\end{abstract}


\end{frontmatter}

\section{Introduction}
\label{S:1}

Friction reduction is a challenge of fundamental importance for energy saving in the automotive industry. The functionality of currently used friction modifiers still needs to be completely understood to design more sustainable lubricant additives with improved performances. Molybdenum dithiocarbamates (MoDTCs) are a class of widely used additives for engine oil capable of reducing friction at the steel-on-steel contact because of the formation of layers of molybdenum disulfide (MoS$_2$)~\cite{yamamotogondo}, which is a well-known solid lubricant. MoDTC complexes, in tribological conditions, undergo molecular dissociation and the originated fragments can recombine to form the slippery tribolayers. The dissociation mechanism of MoDTC is debated, as two different patterns were proposed for the first steps of the reactions, based on experimental observations~\cite{grossiord,Morina1}. Recently, we showed that different dissociation mechanisms are expected for different isomers that can be distinguished by the position of sulfur and oxygen atoms in the complexes~\cite{fe-modtc}. Another relevant aspect in the tribochemistry of MoDTC complexes is the role played by oxygen. De Feo et al. described the quick oxidation of MoDTC-containing samples when heated at 160$\degree$C~\cite{DeFeo15}. The chemical transformations occurring during the oxidative degradation impact the dissociation mechanism of MoDTC and its lubricating properties due to the different morphology and composition of the generated MoS$_2$ films~\cite{flesichauer}. Several studies showed that the presence of zinc dithiophosphates (ZnDTP), common anti-wear additives, is beneficial to achieve low friction coefficients~\cite{Grossiord2000, bec, tan, somayaji2009}. Moreover, interactions between the additives and the substrates play a fundamental role in determining the efficiency of the lubrication~\cite{neville}. When MoDTC is added to the lubricant oil during steel on steel contacts, low friction can be achieved after a run-in period, the duration of which can be influenced by additive-substrate interactions~\cite{grossiord,Morina1}. During scratching, the passivating layers of the metallic surfaces can be removed and the native metal can be exposed~\cite{Clelia2011}, increasing the rate of the tribochemical reactions. In fact, a few authors suggested that this mechanism could initiate the tribochemistry of MoDTC~\cite{graham2001,COUSSEAU2016}. However, Khaemba observed that the formation of MoS$_2$ occurs even in presence of oxidized surfaces~\cite{khaemba}. Here, we elucidate the debated effect of oxygen as a passivating element on iron surfaces in the dissociation of MoDTC. In particular, we compare density functional theory (DFT) and quantum mechanics/molecular mechanics (QM/MM)~\cite{karplus,warshel} simulations, carried out to describe the adsorption of MoDTC on the clean and oxygen-passivated surfaces, with preliminary tribological tests performed on mirror polished and passivated steel discs. The QM/MM approach was in fact shown to be particularly convenient to study such complex reactivity~\cite{fe-modtc}, as it allows to accurately treat the electronic properties of the chemically active species while keeping the computation reasonably affordable~\cite{RestucciaQMMM19,RestucciaQMMM20}. This study aims at clarifying the role of oxygen as a passivating agent in the tribochemistry of MoDTC, especially in the run-in period.

\section{Methods and Systems}

\subsection{Computational Approach}

\begin{figure}[h]
\centering
\includegraphics[width=\linewidth]{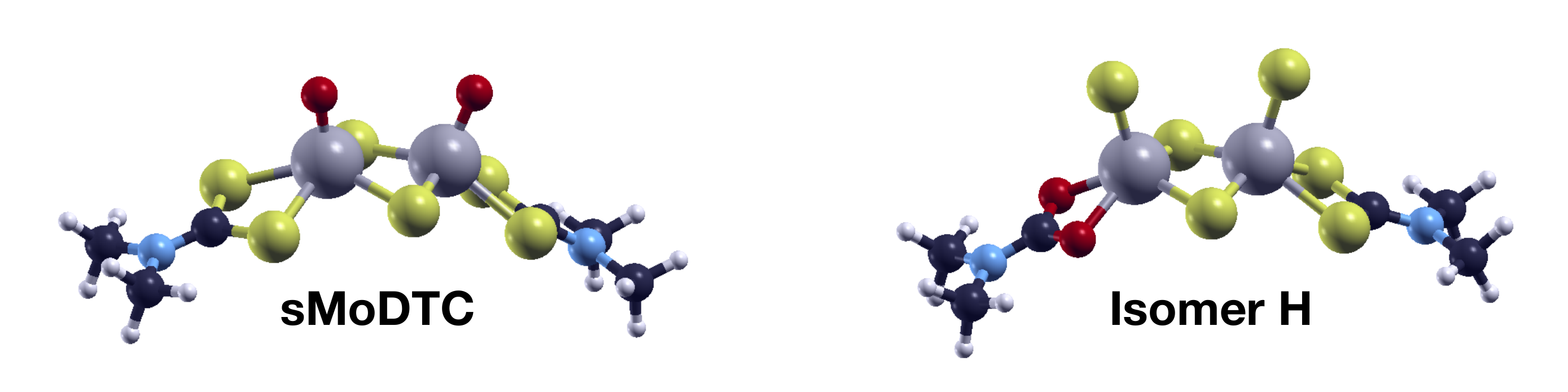}
\caption{Chemical structures of the two complexes considered for the ab initio simulations, namely sMoDTC and Isomer H. The white, black, light blue, red, yellow and grey coloring of the atoms correspond to hydrogen, carbon, nitrogen, oxygen, sulfur and molybdenum atoms, respectively.}\label{fig:complexes}
\end{figure}

We investigated the interaction of the two MoDTC complexes, shown in Figure~\ref{fig:complexes}, with iron. They correspond to the standard MoDTC (sMoDTC) structure and the most stable isomer, named Isomer H in our previous work~\cite{static-modtc}, where the oxygen atoms are moved from the terminal position, on top of molybdenum atoms, to the ligand position, between carbon and molybdenum atoms. We considered the (110) surface of iron, which is the most favorable one for this metal~\cite{giulio}, as a simple approximation for steel. Periodic supercells were employed for all the calculations. In the adsorption calculations, supercells with a $8 \times 2\sqrt{2}$ in-plane size were used, containing a four-layer-thick iron slab and a single MoDTC molecule adsorbed on it. In these calculations, bare and oxidized surfaces were considered, with a coverage by oxygen of 0.25 monolayers and the oxygen atoms occupying the most stable adsorption sites, namely the threefold sites of the Fe(110) surface~\cite{ossowski}.

\begin{figure}[h]
\centering
\includegraphics[scale=0.2]{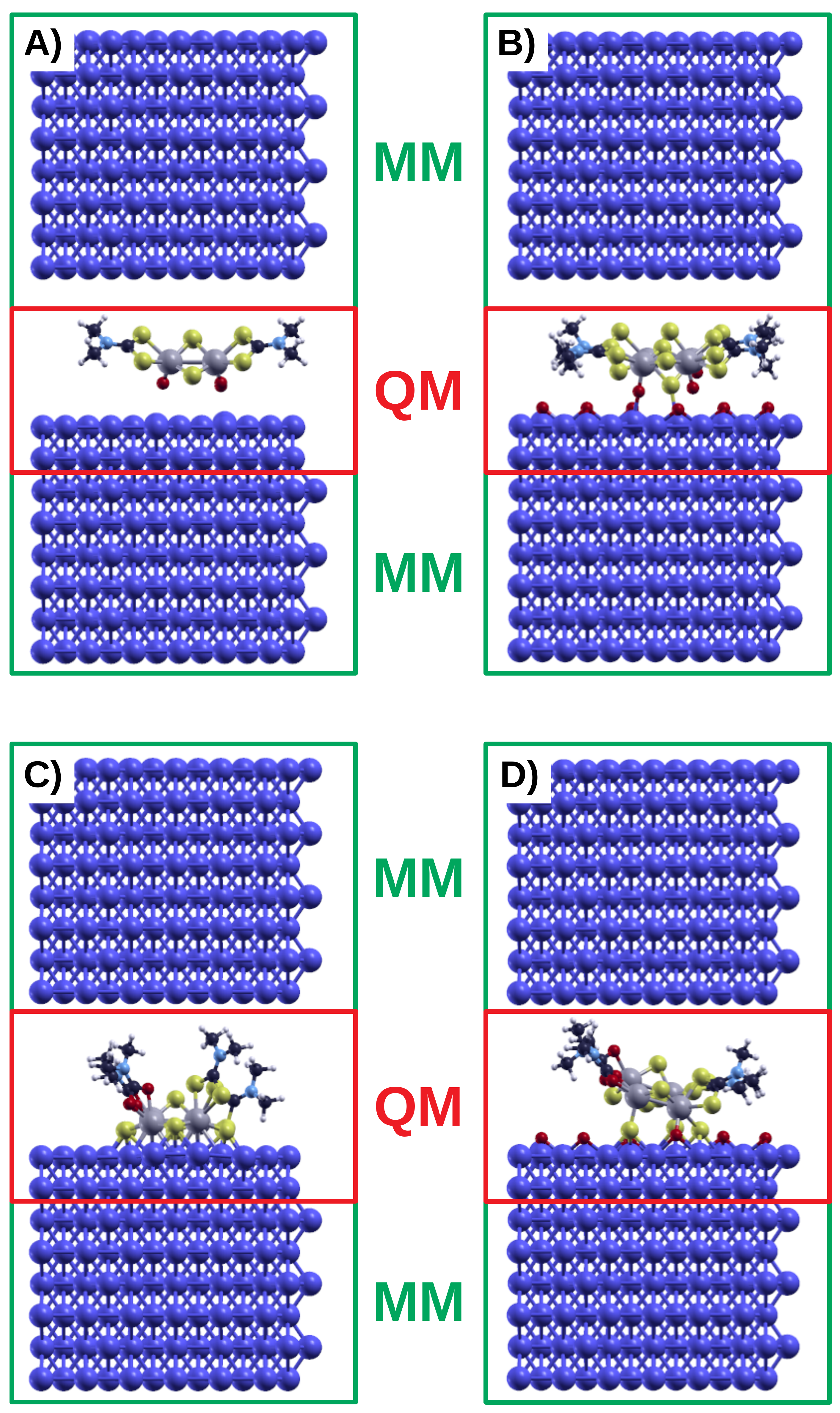}
\caption{Lateral views of the QM/MM systems, with a schematic representation of the partitioning. Initial configurations for sMoDTC and Isomer H on clean (A, C) and passivated (B, D) surfaces, respectively. The blue coloring of the atoms identifies iron atoms.}\label{fig:qmmm-lat}
\end{figure}

In the QM/MM simulations, two MoDTC complexes were put in $6 \times 3\sqrt{2}$ supercells and the top and bottom iron slabs were composed by eight atomic layers each. Both clean and oxygen-passivated surfaces were considered in the dynamic simulations. The two molecules at the interface and the two uppermost iron layers of the bottom slab, with the adsorbed oxygen when present, were part of the chemically reactive region (I region), treated at the DFT level by the Quantum ESPRESSO suite~\cite{QE1}. The remaining six iron layers of the bottom slab and the top iron slab (II region) were treated classically, by using the embedded-atom model~\cite{EAM0,EAM}, with the LAMMPS software~\cite{lammps} driving the whole dynamics. A scheme of the QM/MM simulation setup is shown in Figure~\ref{fig:qmmm-lat}. The atoms of the chemically reactive region were repelled by the top slab as they experienced a purely repulsive Lennard-Jones potential. With this potential, the top slab merely acted as a non-interactive wall on the complexes. The geometry of all the iron layers of the bottom slab, except for the uppermost one, was kept frozen during the dynamic simulation.  As described in a previous publication from our group~\cite{fe-modtc}, the coupling between the two subsystems was realized by employing a subtractive scheme~\cite{senn-thiel}:

\begin{equation}\label{eqn:qmmm}
  E_{tot} = E^\mathrm{I+II}_\mathrm{MM} - E^\mathrm{I}_\mathrm{MM} + E^\mathrm{I}_\mathrm{QM}
\end{equation}

\noindent
where $E_{tot}$ is the total energy of the whole QM/MM system, $E^\mathrm{I+II}_\mathrm{MM}$ is the total energy of the whole system computed with the classical force fields, while $E^\mathrm{I}_\mathrm{MM}$ and $E^\mathrm{I}_\mathrm{QM}$ are the total energies of the reactive region, computed classically and at the quantum level, respectively. An initial thermalization process, lasted one picosecond, allowed the system to globally reach 300 K by employing a Nos\'e-Hoover thermostat with a damping parameter of 33.3 fs. An ensemble of velocities for all the atoms, except for the frozen ones, was generated from a Gaussian distribution with a standard deviation scaled to produce the temperature of 300 K. After the thermalization process, the temperature of the chemically reactive region was left free to evolve and, in order to simulate the sliding of the surfaces, a shear motion of the top iron slab was applied, with velocity kicks of 600 m/s applied at each picosecond on the external layer of the top slab. The timestep of the simulation was 1 fs. The two complexes at the interface experienced a load of 4 GPa applied on the external layer of the top slab for the whole simulation.

The Perdew-Burke-Ernzerhof (PBE)~\cite{PBE} approximation was chosen for the exchange-correlation functional in the DFT calculations. Ultrasoft pseudopotentials were employed and plane waves were used as a basis set for the wave functions. The expansion of the plane waves (charge density) was truncated using a 30 Ry (240 Ry) cutoff for the kinetic energy. The choice for the value of 30 Ry was justified in previous works~\cite{giulio}. Spin polarization was taken into account to allow the spin multiplicity to vary in all the chemical species. We added a Gaussian smearing of 0.02 Rydberg to ease the convergence of the calculations. The evaluation of the charge density was performed at the gamma point in the case of the QM/MM simulations, while in the case of the static calculations a 1x3x1 Monkhorst-Pack grid was employed. We decided not to include any dispersion correction in our calculations because these corrections often overestimate the adsorption energy of the molecules on metallic substrates~\cite{vdw1,vdw2}. The geometry optimizations were stopped when the total energy and the forces converged under the thresholds of $1\cdot10^{-4}$ Ry and $1\cdot10^{-3}$ Ry/bohr, respectively.

\subsection{Experimental Techniques}

To support the evidence provided by the simulations, we carried out an experimental investigation to verify whether an increase in the amount of oxygen on the surface of the metallic substrates can influence the tribological behavior of MoDTC. Mirror-polished discs made of AISI 52100 hardened steel (HRC between 58 and 66), with a diameter of 60 mm and a surface roughness of 0.05 $\mu$m, were chosen because of their higher surface homogeneity, which limits the run-in period, avoiding undesired wear and allowing to focus primarily on the tribochemical activation of MoDTC. All the steel discs were initially cleaned with heptane. The surface of the reference discs consisted of bare steel, whereas a group of discs was heated in air at 170\degree C for 7 h on a temperature-controlled hot plate to promote the exchange of oxygen with the environment. The surface and the lateral profile of the reference and the oxidized discs were analyzed by an FEI Quanta FEG 250 scanning electron microscope coupled to an Oxford X-Max 80 SDD energy-dispersive X-ray spectroscope (SEM-EDX) in approximately 6 mbar of vacuum with a beam of 20 kV and working distance of 1 cm. After the surface analysis, tribological tests were carried out by using an Anton Paar MCR302 rheometer with a T-PID/44 tribology assembly (three balls on disk tribology measurement cell). The diameter of the balls was 6 mm. Each run lasted for 20 minutes, with an exponentially increasing rotation speed from 4.7 mm/s to 377 mm/s over 750 s, an operational temperature of 110\degree C and a contact pressure of 700 MPa. The exponential speed profile allows to study in detail the delayed activation of the additives while always maintaining the boundary lubrication regime. The lubricant consisted of a mixture of MoDTC complexes, Adeka Sakuralube 525, in a group III mineral oil, Yubase 4, with a resulting concentration in molybdenum of about 500 ppm, and was deposited directly on the disc immediately before the run.

\section{Results and Discussion}

\subsection{Molecular Adsorption and Dissociation in Static Conditions}

\begin{figure}[h]
\centering
\includegraphics[width=0.7\linewidth]{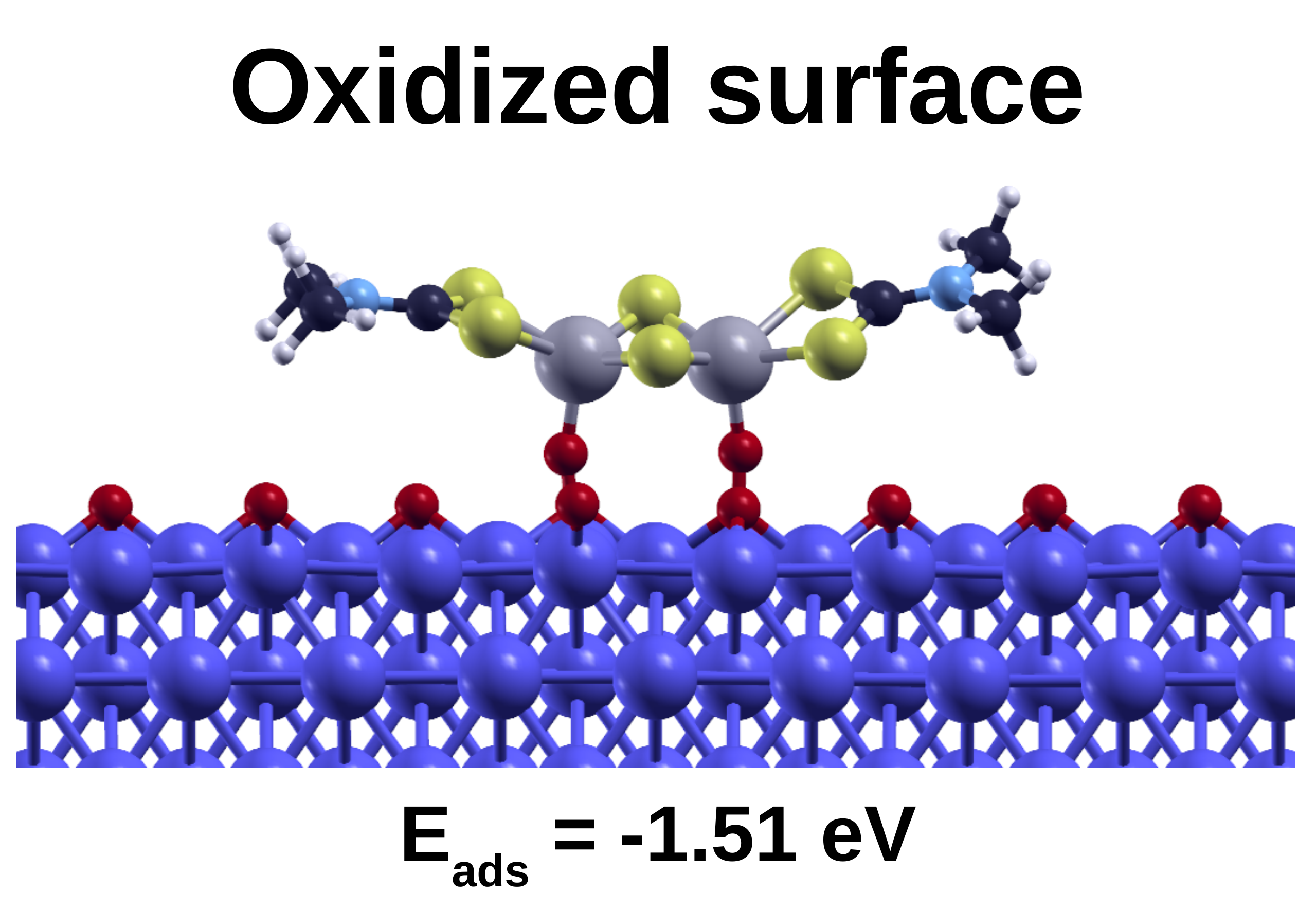}
\caption{Final configuration of the adsorbed sMoDTC complexes on the oxidized surface.}\label{fig:ads}
\end{figure}

The interaction of the MoDTC compounds with the bare metallic substrate favors their molecular dissociation. Initially, MoDTC must adsorb onto the metallic surface for the dissociation to occur, and passivating agents such as oxygen and sulfur may hinder the adsorption of the additives on iron. To investigate this mechanism, we positioned the sMoDTC molecule on the clean and oxidized surfaces and optimized all the degrees of freedom. Indeed, we observed that oxygen on the iron surface limit the adsorption of sMoDTC, as shown in Figure~\ref{fig:ads}. Only the oxygen atoms in terminal position of sMoDTC, pointing in the direction of the surface, are bound to the substrate. All the other atoms are repelled. This behavior is reflected on the adsorption energies $E_{ads}$, calculated as:

\begin{equation}\label{eqn:ads}
  E_{ads} = E_\mathrm{tot} - E_\mathrm{mol} + E_\mathrm{sub}
\end{equation}
\noindent
where $E_\mathrm{tot}$, $E_\mathrm{mol}$ and $E_\mathrm{sub}$ are the total energies of the whole system, of the complex alone and of the substrate alone in the same simulation cell, respectively. In fact, the adsorption energies of sMoDTC on the oxidized surface in the particular configurations shown in Figure~\ref{fig:ads} is -1.51 eV, compared to -4.33 eV in the case of a clean iron surface~\cite{fe-modtc}.

\begin{figure}[h]
\centering
\includegraphics[width=\linewidth]{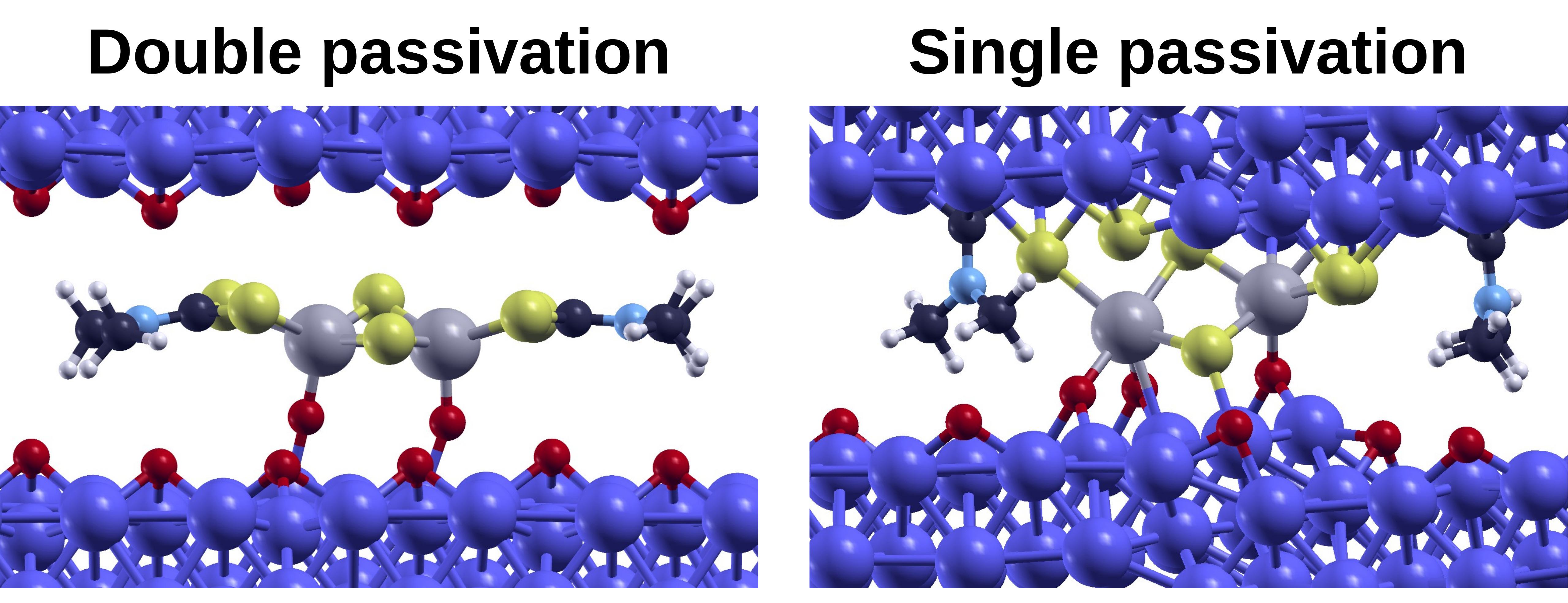}
\caption{Final configurations of the variable-cell calculations with double and single passivating oxygen layers at the iron interface.}\label{fig:vcr}
\end{figure}

Since the adsorption of MoDTC is limited by the presence of oxygen on the iron surface, the catalytic effect of the substrate in promoting the dissociation of the complexes is decreased as well. Variable-cell calculations in which the molecules withstand a load of 0.5 GPa show that the complexes cannot undergo dissociation when the whole interface is passivated. When only one of the two surfaces is covered, the dissociation of the complexes can occur, due to the interaction of the complexes with the only clean surface, as shown in Figure~\ref{fig:vcr}.

\subsection{Molecular Adsorption and Dissociation in Tribological Conditions}
\begin{figure}[h]
\centering
\includegraphics[width=\linewidth]{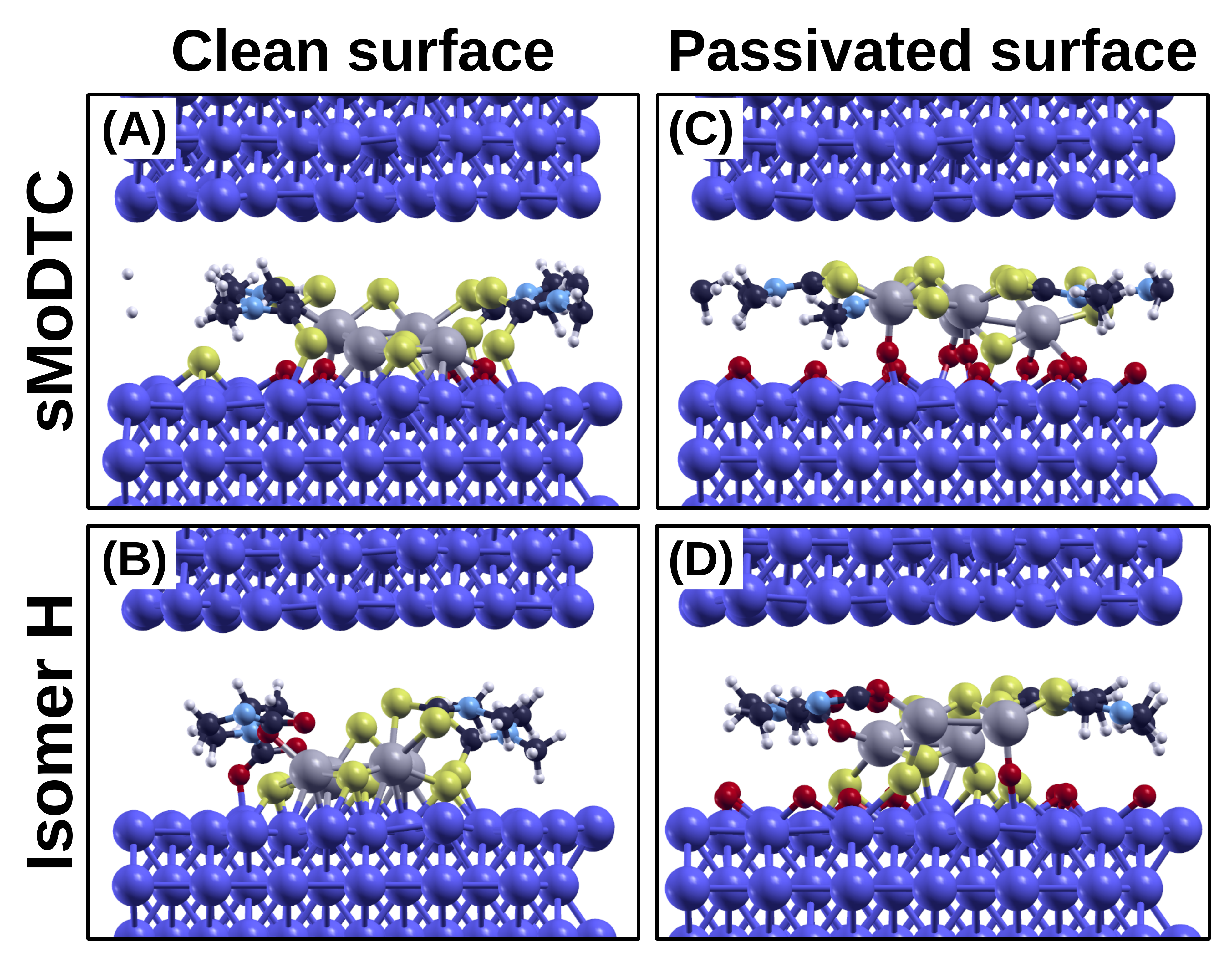}
\caption{Lateral views of the QM/MM systems after 2 ps of the simulation.}\label{fig:qmmm-top}
\end{figure}

Dynamic simulations offer particularly valuable insights into the mechanism of tribochemical reactions at the atomistic level. By following the QM/MM approach, it is possible to accurately monitor the formation and the dissociation of chemical bonds in the first picoseconds of the reactions while maintaining the simulation computationally affordable. For these simulations, a suitable geometry of the systems was chosen to study the effects of the different substrates. The same counter-surface was chosen for all the simulations, in analogy to the experiments, and it was represented simply by a repulsive wall to minimize its chemical effect on the species at the interface.
Lateral views of the QM/MM systems at 2 ps of the simulation are shown in Figure~\ref{fig:qmmm-top}. Both complexes are forced to adsorb onto the bottom iron slabs under the effect of load. The dynamic simulations, in agreement with the static calculations, show that the clean surfaces allow the complexes to chemisorb (Panels A and B). Some of the atoms in the ligand position of sMoDTC and Isomer H are close to the surface, where the catalytic effect of the metallic atoms may induce a dissociation according to the different patterns shown in our previous work~\cite{fe-modtc}: some of the sulfur atoms in the ligand position of sMoDTC detach and occupy free sites on the iron surface (Panel A). On the other hand, the oxygen atoms in the ligand position of Isomer H remain attached to the carbon atom of the carbamate unit (Panel B). When the iron surface is bare, the different decomposition mechanisms of the two isomers can be observed. 

The passivation of iron by oxygen hinders the adsorption of the central units of MoDTC (Panels C, D). In fact, the complexes become flat due to the applied load. The dissociation appears to be less probable in this case, as the metallic substrate cannot efficiently catalyze the chemical reaction. The decreased number of free sites on the surface limits the stabilization effect of the fragments that can be originated upon molecular dissociation, regardless of the presence of oxygen or sulfur in the ligand position of the complexes. The experiments described in the following section demonstrate the consequence of this fundamental mechanism.

\subsection{Characterization and Tribological Tests on the Oxidized Surfaces}

\begin{figure}[h]
\centering
\includegraphics[width=0.8\linewidth]{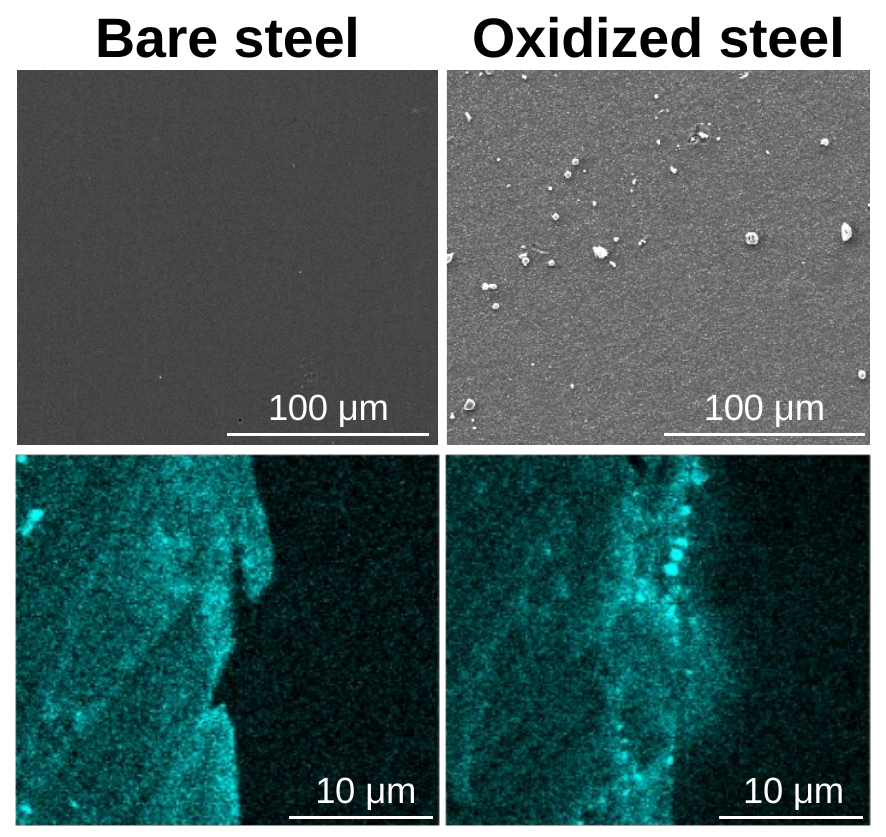}
\caption{Surface analysis of the reference and oxidized steel discs. Top: SEM images of the bare (left) and oxidized (right) surfaces. Bottom: lateral profile of the bare (left) and oxidized (right) surfaces, respectively. The light blue coloring identifies the intensity of the oxygen K$\alpha$1 peak.}\label{fig:sem-edx}
\end{figure}

\begin{figure}[h]
\centering
\includegraphics[width=\linewidth]{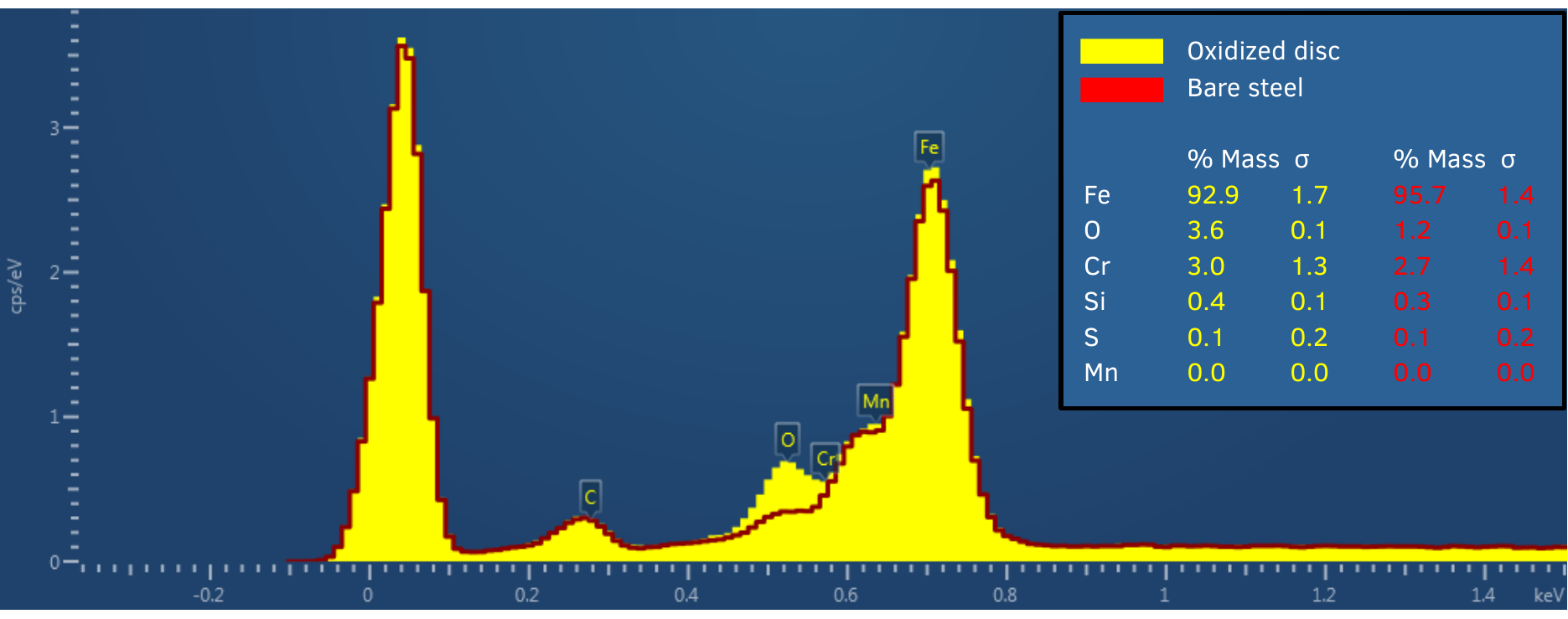}
\caption{Comparison of the quantitative analysis by EDX of the oxidized disc with respect to the bare steel disc.}\label{fig:edx}
\end{figure}

Figure~\ref{fig:sem-edx} reports the results of the surface analysis of the bare and the oxidized surfaces used in the tribological tests. The two different surfaces can be distinguished both with and without microscopy. To the naked eye, the oxidized disc presents a darker color, as shown in the top panels of Figure~\ref{fig:sem-edx}, while maintaining the capability to reflect light. The lateral profile of these surfaces reveals that in both the bare and the oxidized surfaces, the oxygen concentration is higher in the proximity of the surface, as shown in the bottom panels. However, in the case of the oxidized disc, the amount of oxygen in bulk decreases quickly with respect to the surface. Because the two discs were originally identical, this result suggests that the increase in the surface concentration of oxygen was effectively achieved. The quantitative analysis of the EDX spectrum shown in Figure~\ref{fig:edx} reveals, in fact, that the mass percentage of oxygen on the surface becomes three times higher with respect to the reference disc, as expected by the preparation protocol

It is hard to control the thickness and the stoichiometry of the iron oxides with this heating treatment. Therefore, Fe$_2$O$_3$ and Fe$_3$O$_4$ with different surface orientations might be present at the same time on these samples. Previous computational investigations indicate that the surface energies of the different orientations of Fe$_2$O$_3$ and Fe$_3$O$_4$ are comparable, with the lowest and the highest being 0.96 and 1.37 J/m$^2$ for the (001) surface of Fe$_3$O$_4$ and the (100) surface of Fe$_2$O$_3$, respectively~\cite{fe3o4,fe2o3,ironoxides}. Therefore, we do not expect significant differences in the reactivity of the MoDTC compounds based on the stoichiometry of the oxide layer. The calculated surface energy of iron (110) is approximately 2.5 J/m$^2$~\cite{Wolloch2019} and explains the enhanced reactivity of the compounds on the bare metal. Even though the surface energy for the experimental samples was not directly measured, it is reasonable to expect a higher energy value for bare steel with respect to oxidized steel.

\begin{figure}[h]
\centering
\includegraphics[width=0.75\linewidth]{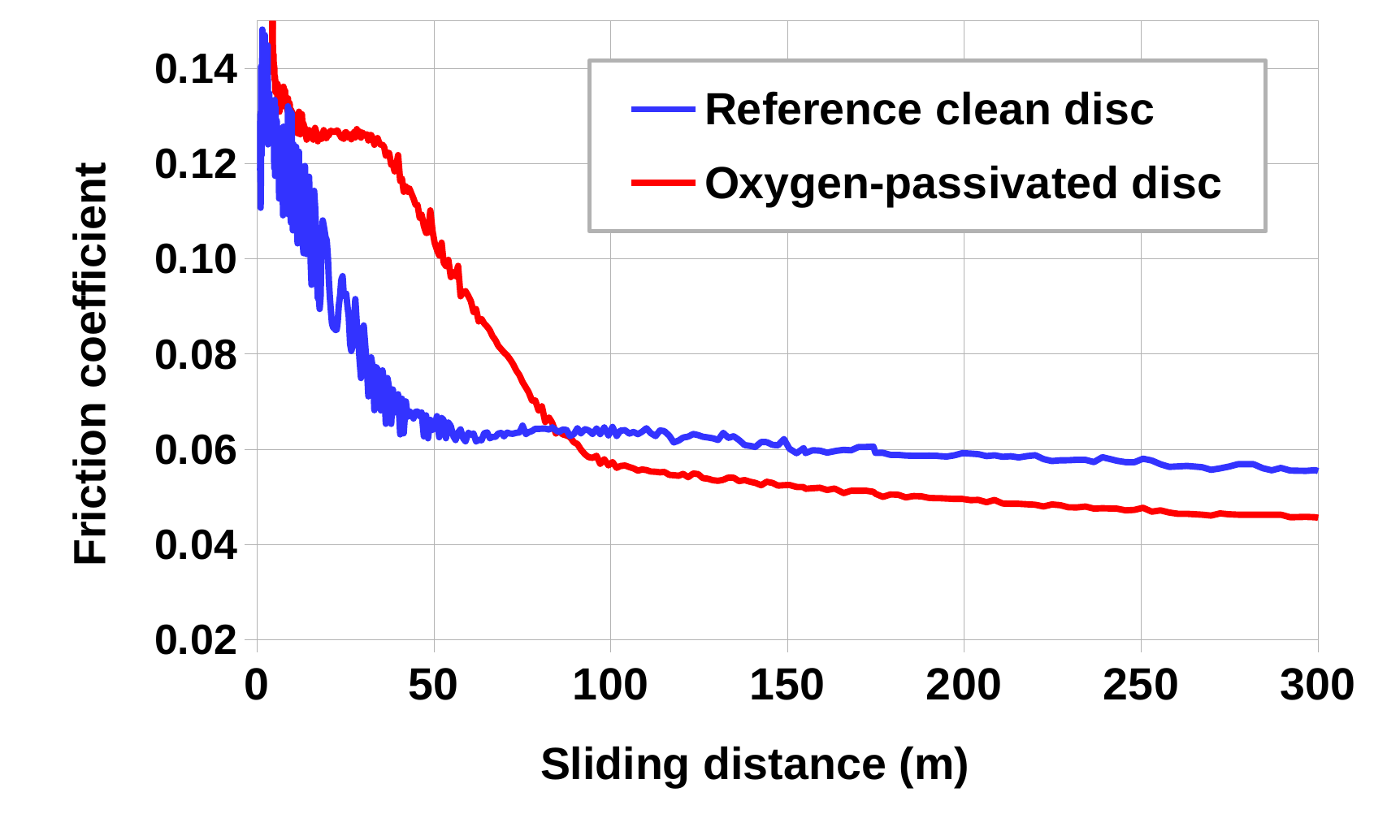}
\caption{Tribological tests of the oxidized and the clean mirror discs under lubrication of S525.}\label{fig:tt}
\end{figure}

The tribological tests carried out on these surfaces, shown in Figure~\ref{fig:tt}, reveal that the surface oxidation leads to high friction during the run-in period compared to the reference disc. After the first 180 seconds of the run, at an approximate sliding distance of 8.7 m, the activation of the MoDTC additives occurs and friction drops to values comparable to the ones of the reference disc, as a sufficient amount of MoS$_2$ is generated to maintain low friction. Furthermore, the friction coefficient of the oxidized surface becomes even slightly lower than the reference at the end of the run. The high friction in the early stages of the run might be associated to the removal of oxidized material which is not able to fully interact with the MoDTC compounds. After reaching more reactive layers below the oxidized surface, the substrate behaves as it was not oxidized by thermal treatment, and the energy excess due to the initial catastrophic events may contribute to an enhanced reactivity of the MoDTC complexes, leading to an even lower friction coefficient. For the oxidized steel discs, the quick formation of debris caused by local stresses at the asperity contact during the initial stage of the run increases the possibility for the additives to encounter reactive metallic sites. These sites can catalyze the tribochemical reactions and, therefore, promote the dissociation of MoDTC or induce isomerizations and S-O substitutions in these additives. As explained more in detail in the following Section, the chemistry of the MoDTC additives is relevant because different compounds undergo different decomposition paths on the metallic substrates.

\subsection{Discussion}
The ab initio calculations showed that an iron surface free of adsorbates is beneficial to chemisorb MoDTC compounds and that the chemisorption of MoDTC is a necessary step in the formation of MoS$_2$. The dissociation of MoDTC compounds can be activated thermally even in the absence of mechanical stresses. However, shear stress has been demonstrated to promote dissociation at relatively low temperatures~\cite{khaemba}. The synergy of the extreme conditions observed during sliding is translated into quick run-in periods, as a sufficient amount of MoS$_2$ is immediately generated to achieve low friction~\cite{graham2001}, as observed in the case of the reference clean discs used in this work. The fact that MoDTC cannot access the catalytic sites of the substrate when the metal is oxidized slows down the production of MoS$_2$, delaying the typical friction drop. Such a mechanism would match the observations by Cousseau et al.~\cite{COUSSEAU2016}, who suggest that the removal of the oxidized layers of the substrate, because of the contact of the asperities, is beneficial to promote the transformation of MoDTC into MoS$_2$.

Khaemba observed that MoS$_2$ can be formed even when the surface is oxidized~\cite{khaemba}. However, the resulting tribolayer is less durable, because it cannot adhere strongly to the substrate. Although our investigation did not focus on the durability of the MoS$_2$ tribolayers, our simulations can explain these experimental observations. The adsorption of the MoDTC structures is limited by the presence of the oxygen atoms in all of our calculations. Low durability of the MoS$_2$ tribolayer can be expected due to such a limited adsorption of the additives, when the surface is rich in oxygen. Even though the low MoDTC concentrations we considered in the simulation cells do not allow the formation of complete tribolayers, the formation of MoS$_2$ can be expected in all cases, provided that a sufficient amount of molecules is present at the contact. The MoS$_2$ layer can be strongly bound to the surface in the case of bare steel, whereas the MoDTC compounds should cluster and recombine while being only weakly adsorbed in the case of the oxidized surface.

\begin{figure}[h]
\centering
\includegraphics[width=0.75\linewidth]{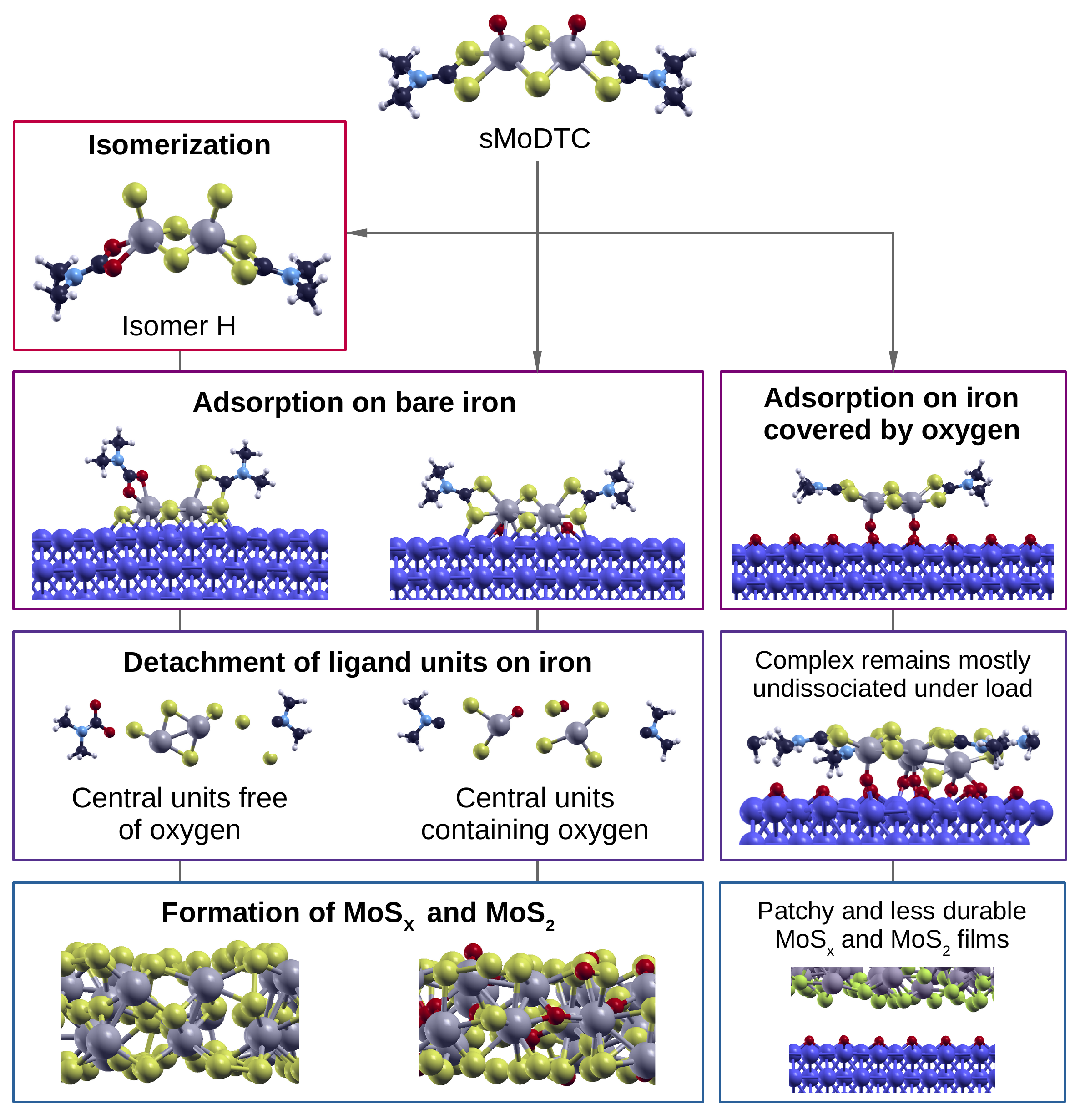}
\caption{Mechanism of the dissociation of MoDTC complexes on iron. Left side: when the metallic substrate is clean, the different chemistry of the additives can lead to tribofilms with different properties. Right side: when the surface is covered by oxygen, the reactivity of the complexes is generally hindered.}\label{fig:steps}
\end{figure}

Based on the observations derived from our previous work~\cite{fe-modtc} and the new evidence from this study, we propose a reaction mechanism for the MoDTC complexes that is schematically represented in Figure~\ref{fig:steps}. The standard MoDTC complex or its isomers in the lubricant mixture can come in contact with the metallic substrate. The most stable isomer of MoDTC is Isomer H, the complex with two oxygen atoms in ligand position, as demonstrated by previous DFT calculations~\cite{static-modtc} and mass spectrometry performed on aged samples containing MoDTC~\cite{DeFeo15}. When the iron substrate is bare, any MoDTC complex can chemisorb onto the metallic surface. The presence of mechanical stresses promotes the decomposition of the complexes, and the resulting molecular fragments depend on the presence of oxygen atoms in the ligand position. MoDTC structures with oxygen atoms in ligand position release central units ideal to form MoS$_2$ because they contain only molybdenum and sulfur. In the case of sMoDTC, the oxygen atoms in terminal position remain close to the Mo atoms and become part of the MoS$_x$ network that will eventually become MoS$_2$. This mechanism is different when the surface is covered by oxygen. Indeed, the MoDTC complex weakly physisorbs on the oxidized substrate and the decomposition is significantly hindered even in the presence of mechanical stresses, regardless of the position of the oxygen atoms in the complex. When the surface is oxidized, the only possibility to form MoS$_x$ is when MoDTC structures interact and form clusters containing Mo and S atoms without the catalytic mediation of the metallic substrate. This leads to inhomogeneous and unstable MoS$_2$ tribofilms.

\section{Conclusions}

Based on the results obtained by means of tribological experiments and ab initio simulations, we showed that the oxidation of the metallic surface hinders the tribochemical reactivity of MoDTC lubricant additives both in static and tribological conditions. Direct comparison between the simulations and the real tribological interface is, in general, a formidable task, and searching for an exact match between the conditions of the two approaches lies beyond the purpose of this work. A more detailed study would be required to fully describe the tribological behavior of MoDTC on the oxidized steel samples. However, the qualitative observations obtained from our preliminary tribological tests and the results offered by the simulations matched. An increased amount of oxygen on the surface of the mirror-polished steel discs showcased increased friction in the initial phase of the tribological tests, meaning that the functionality of the MoDTC lubricant additives was inefficient and resulted in a much longer run-in period. The increased friction in the case of the oxidized surface induced harder tribological conditions on the additives and possibly higher levels of oxidation of MoDTC. We showed in a previous publication that this phenomenon can favor the release of molecular units ready to build MoS$_2$, leading to an even lower friction coefficient in the end. The quantum mechanics/molecular mechanics simulations demonstrated in fact that the presence oxygen as a passivating species limits the possibility of MoDTC to access the catalytic sites of the metal, reducing its capability to dissociate at the beginning of the sliding, regardless of the distribution of the oxygen atoms in the molecules. Indeed, the different decomposition mechanism previously described for sMoDTC and Isomer H cannot be clearly observed when the metallic surfaces are oxidized. These findings are
key to formulate a complete description of the functionality of these lubricant
additives, demonstrating the power of ab initio simulations combined with
experiments in clarifying the complex tribochemistry of friction modifiers.

\section*{Acknowledgments}

We are thankful to Quentin Arnoux for his contribution in the experimental characterization of the steel discs. Several pictures in this work were created with the help of XCrySDen~\cite{xcrysden}.

\section*{Funding}
These results are part of the “Advancing Solid Interface and Lubricants by First-Principles Material Design (SLIDE)” project that has received funding from the European Research Council (ERC) under the European Union’s Horizon 2020 research and innovation program (Grant Agreement No. 865633).



\bibliographystyle{elsarticle-num-names}
\bibliography{bibliography.bib}


\end{document}